# Opto-magnetic imaging of neural network activity in brain slices at high resolution using color centers in diamond

Mürsel Karadas[1], Adam M. Wojciechowski[2], Alexander Huck[2], Nils Ole Dalby[2,3], Ulrik Lund Andersen[2], Axel Thielscher[1,4]

**Institutions:**

[1]Department of Electrical Engineering, Technical University of Denmark, 2800 Kongens Lyngby, Denmark

[2]Department of Physics, Technical University of Denmark, 2800 Kongens Lyngby, Denmark

[3]Department of Drug Design and Pharmacology, Copenhagen University, 2100 Copenhagen, Denmark

[4]Danish Research Center for Magnetic Resonance, Copenhagen University Hospital, 2650 Hvidovre, Denmark

**Correspondence by:**

Dr. Axel Thielscher
Danish Research Center for Magnetic Resonance
Copenhagen University Hospital Hvidovre
2650 Hvidovre, Denmark
Email: axelt@drcmr.dk
Phone #: 0045-38623326

## Abstract

We suggest a novel approach for wide-field imaging of the neural network dynamics of brain slices that uses highly sensitivity magnetometry based on nitrogen-vacancy (NV) centers in diamond. In-vitro recordings in brain slices is a proven method for the characterization of electrical neural activity and has strongly contributed to our understanding of the mechanisms that govern neural information processing. However, traditional recordings can only acquire signals from a few positions simultaneously, which severely limits their ability to characterize the dynamics of the underlying neural networks. We suggest to radically extend the scope of this method using the wide-field imaging of the neural magnetic fields across the slice by means of NV magnetometry. Employing comprehensive computational simulations and theoretical analyses, we characterize the spatiotemporal characteristics of the neural magnetic fields and derive the required key performance parameters of an imaging setup based on NV magnetometry. In particular, we determine how the technical parameters determine the achievable spatial resolution for an optimal reconstruction of the neural currents from the measured field distributions. Finally, we compare the imaging of neural slice activity with that of a single planar pyramidal cell. Our results suggest that imaging of neural slice activity will be possible with the upcoming generation of NV magnetic field sensors, while imaging of the activity of a single planar cell remains more challenging.

# Introduction

Nitrogen-vacancy (NV) color centers in diamond are currently emerging as a novel and highly sensitive tool to measure magnetic fields at ambient temperatures[1–3]. The measurement is based on optically detected magnetic resonance[4], where the signal is obtained through a measurement of the NV fluorescence level induced by an external magnetic field. This technique opens up new venues to measure weak magnetic fields occurring in biological systems [5,6], and might particularly useful to characterize neural activity[7]. When combined with wide-field imaging sensors to read out the florescence at high spatial and temporal resolution, it might offer the possibility to image the dynamics of neural networks in great detail (Fig. 1A&B).

In-vitro recordings in thin brain slices have been a mainstay in the repertoire of electrophysiological methods for the recording of neural activity. Since it was demonstrated that in brain slices similar activity could be recorded as in intact animals[8], this approach has been used as a test bed to decipher the fundamental mechanisms that govern neural interactions. Using careful cutting techniques, intact neural networks can be maintained and studied, while at the same time a good accessibility of specific neurons and neural pathways is obtained. However, the traditional method is severely limited by the low number of electrodes which can be used for simultaneous recordings. While detailed information is gained from a few slice positions, an overview of the complete network dynamics is lacking.

In this article, we suggest to radically extend the scope of slice recordings by combining them with the wide-field imaging of neural activity via NV centers in diamonds. Adding simultaneous high-resolution recordings from a slice area of 1x1 mm² or more would give detailed access to the neural dynamics across extended local networks, as for example occurring in the hippocampus region of rat or mice brains[9] (Fig. 1A). Our analysis is based on detailed computer simulations and demonstrates the feasibility of this approach. We focus on a scenario in which magnetic fields are generated by synchronous neural activity in a subpart of the hippocampal CA1 region in a mouse brain slice, elicited by electrical stimulation of the Schaffer collaterals projecting from CA3 to CA1[10] (Fig. 1A&B). We chose this scenario, because (i) there is a comprehensive collection of data related to CA1 neurons and their models are validated by numerous experiments [11], (ii) the topology of CA1 is well known, (iii) it is easy to obtain field recordings and intracellular recordings due to the regular formation of the cells, (iv) the Schaffer collateral axons form a homogeneous pathway that is easily activated to study synaptic transmission and plasticity, and (v) it is easy to keep the cells in this region alive and functional during the slice preparation. The dynamic response of the CA1 pyramidal cells to the stimulation are modelled in NEURON [12], using previously reported morphology and biophysical properties [13]. We then determine the extracellular magnetic fields at the diamond sensor surface and for comparison also the electric potential. These quantities are obtained from the simulated membrane potentials and transmembrane currents using a forward modelling scheme (see Methods and Supplementary Material). Based on comprehensive simulation results, we conclude on the expected magnetic field strengths created by the evoked neural activity and the required temporal resolution of the recording system. We further simulate with a Wiener deconvolution filter approach the optimal reconstruction of the spatial distribution of the neural currents. This allows us to systematically determine the achievable spatial resolution in dependence on two key technical parameters of the system, the in-plane resolution of the sensor and its noise level.

In addition to our main scenario, we re-evaluate the simulations of the magnetic fields of a single planar CA1 pyramidal cell (Fig. 2A). We find diverging results to those originally reported [14] with lower expected field strengths and suggest putative causes for the observed discrepancies.

## Results

*Planar CA1 Cell*

We begin our analysis by re-evaluating a single planar CA1 pyramidal cell [14–16] (Fig. 2A). To simulate identical cases as tested in a prior study [14], the neuron is stimulated with a 10 ms long injected current of 2 nA at 21 distal dendrites. The resulting transmembrane potential at the soma for different ambient temperatures is plotted in Fig. 2B. As expected, the duration of the action potential increases with decreasing temperature, but only for temperatures much below ~20°C similar durations are obtained that resemble the previously reported ones [14]. The resulting peak magnetic field distribution is shown in Fig. 2C, obtained for an ambient temperature of 25°C. Compared to the original results [14], our simulations suggest nearly 4-5 times smaller peak magnetic field strengths, which has several possible reasons: First, the assumed distance of 100 nm between sensor surface and the cell should be measured relative to the lower surface of the cell structure as the cell lies on the surface. In fact, we obtain similar values for the field strengths as originally reported, when we measure the distance relative to the center of the cell structure. This is physically not possible, as then parts of the cells would be inside the sensor. Second, in contrast to our forward modeling scheme, the scheme used in the prior study fits the time course of the transmembrane potential by a sum of Gaussian functions. This can result in fitting errors that may contribute to the observed discrepancies. Third, the value of the ambient temperature was not mentioned in the prior study, but has a strong influence on the speed at which action potentials develop. In that study, the time courses reported for the transmembrane potential are very slow and could not accurately be replicated in our simulations even when assuming an ambient temperature of 10°C. However, neither the second nor the third reason can fully explain the differences in the reported magnetic field strengths.

*Hippocampus Slice: Spatiotemporal distribution of the neural magnetic activity*

As main scenario, we assess the stimulation of a CA1 subarea with an assumed active size of 500 µm x 500 µm x 300 µm (width x length x height) placed centrally on a diamond sensor (Fig. 1B). In order to account for dysfunctional cells caused by the preparation procedure, we add 50 µm thick subareas each on top and below the active CA1 volume, yielding a total slice thickness of 400 µm. A temperature of 35°C is chosen. These values are within the standard thickness and temperature ranges for hippocampus slices used in electrophysiological experiments.

The magnetic field and the electric potential caused by the neural activity is calculated at a diamond surface with a size of 1x1 mm². Unless indicated differently, the surface is discretized using a 20x20 grid, resulting in an in-plane resolution of 50 µm. We simulated the electrical stimulation of the Schaffer collaterals using a temporally synchronous excitation of the excitatory AMPA synapses of the pyramidal cells, which are situated on the apical dendrites (the stratum radiatum, S.R.; Fig. 1A) and the basal dendrites (the stratum oriens, S.O.)[17]. The synaptic events occurred at two successive time points (set to 12.5 ms and 37.5 ms) to mimic a repeated electric stimulation at 40 Hz (shown in the upper rows of Fig 5). Additionally, we included a jittering in our model using a normal distribution with standard deviation $\sigma$ ($\sigma = 0.25$, unless indicated differently), which mimics the jitter in the excitatory inputs that might occur due to slightly different path delays. We also varied the synaptic strength in order to distinguish between a "non-spiking" case and a "spiking" case. In the non-spiking case, a strength of 0.3 nS was empirically selected that was just low enough to avoid action potentials in order to mimic strong sub-threshold activity. In the spiking case, a strength of 0.6 nS was chosen that was just high enough to reliably induce action potentials in all model neurons. Further details of the neural dynamics are stated in the Supplementary Material.

The spatial distribution of the local field potentials (LFP) and the peak magnetic field strength for stimulation of the CA1 subarea are shown in Fig. 3. For both the spiking (Fig. 3A) and the non-spiking case (Fig. 3D), the LFPs show the expected spatial patterns[18–20] and their strength is in accordance with previous studies. The $B_Y$ component of the magnetic field is negligibly small in both cases (data not shown), in accordance with the orientation of the pyramidal cells determining the neural source currents to be mainly in the Y direction. For the spiking case, the magnitude of the magnetic field $B_X$ component reaches peak values of up to 2.5 nT in the sensor plane at positions directly beneath the stimulated subarea (Fig. 3B). The $B_Z$ component reaches its peak values of around 1 nT at the left and right hand side of the subarea (Fig. 3C) and the spatial patterns of both the $B_X$ and $B_Z$ component are in accordance with source currents that mainly flow along the Y direction. In the non-spiking case (Figs. 3E&F), the peak strengths of the $B_X$ component is reduced to around 0.36 nT, and the peak strength of the $B_Z$ component is around 0.17 nT.

We also simulated the spatiotemporal field distributions that would occur when stimulation of the Schaffer collateral only activated the S.R. region of the pyramidal cells (Fig. S4). The peak values of the LFP and the magnetic field remain in the same range, but the spatial distributions clearly changes. This indicates that both the electric and magnetic recordings reveal detailed information about the spatiotemporal distribution of neural activity in the slice.

*Effects of the width and thickness of the activated CA1 subarea on the peak magnetic fields*

Next, we characterized the dependence of the expected peak magnetic fields on the thickness of the hippocampus slice and the width of the activated subarea. When systematically increasing the width of the active region (X direction in Fig. 1B) while keeping a constant slice thickness of 300 µm, the peak magnetic field strength will start to saturate at a width of ~300 µm (Fig. 4A&C). Similarly, systematically increasing the slice thickness (Z direction in Fig. 1B) for a fixed width of 500 µm results in increasing peak magnetic fields that begin to saturate for a slice thickness of ~300 µm. For even thicker slices, the upper parts of the slice will not contribute markedly to the field measured by the sensor due to the steep decay of the magnetic field strength with distance to the neural sources.

The saturation is more clearly visible for the spiking vs. the non-spiking case. The reason is that the axially oriented current distributions that occur during the action potentials have the form of dipoles (Fig. 3B), causing a steeper decay in the field strength with increasing distance to the measurement site.

*Effect of synaptic strength and temporal synchrony on the magnetic signal strength*

In the above simulations of the spiking case, the synaptic strengths were chosen just high enough to activate all pyramidal cells in the subarea and the temporal jitter was chosen such that the duration of population spike in the LFPs was in the range of those seen in physiological slice recordings[21]. In the following, we aim to ensure the robustness of our results by testing how much the simulated peak magnetic fields depend on these choices. For completeness, we also report on the impact of the temporal jitter on the magnetic fields for the non-spiking case. In order to speed up the calculations and without loss of generality, these simulations were obtained for a small region containing 200 cells. While the peak field strength increases only moderately for highly synchronous input (second rows of Figs. 5A vs. 5B), it decreases around twofold when the input is less synchronous (second rows of Fig. 5C vs. 5B). Doubling the synaptic strength from 0.6 nS to 1.2 nS increases the peak magnetic fields for the spiking case in the range 33%-50%, unless the synaptic events are highly asynchronous. Taken together, given the conservative estimates of the temporal synchrony and synaptic strength used for the spiking case in the main part of the results, in an experiment we might expect moderately higher peak magnetic field strengths than estimated here.

Given that the reported magnetic fields strongly depend on the simulated neural axial current densities, we in addition validated our simulations by determining an equivalent current dipole (ECD) by summing the simulated neural currents and comparing the simulated ECD strength with values reported for hippocampal pyramidal cells[22]. The ECD is given by $\vec{Q_i} = \sum_k I_i^k \vec{L}^k$, with $L^k$ indicating the length of the $k^{th}$ cylinder and $I_i^k$ the axial current flow in the cylinder k at the time frame i [22]. Previously the ECD of a single cell is estimated as 0.2 pA-m when an AP occurs[8]. Accordingly, in our simulation of 200 highly synchronized neurons (Fig. 5A), in which the ECDs of the single neurons are simultaneous enough to add up linearly, the total ECD reaches up to 35 pA-m. The remaining small discrepancy may, e.g. result from slightly different cell morphologies.

*Temporal bandwidth of the neural magnetic fields*

For the spiking case, most of the signal power is still maintained when using a cut-off frequency of 400 Hz (Fig. 6A&C), suggesting a temporal sampling rate of 800 Hz according to the Nyquist criterion. For the non-spiking case, a cut-off frequency of 150 Hz still maintains most of the signal power (Fig. 6B&D), indicating that a sampling rate of 300 Hz is sufficient.

*Achievable spatial resolution and required in-plane resolution and sensitivity of the sensor*

In order to characterize the achievable spatial resolution and its dependence on the system parameters, we simulated the magnetic field of a point-like axial current density and reconstructed its 2D projection via an optimal Wiener deconvolution filter (see Methods). A schematic of the signal analysis pathway is presented in Fig. 7A. We report the point spread function of the reconstructed signal to characterize the achievable spatial resolution. In addition, we report the peak signal to noise ratio (pSNR)[23] of the reconstructed signal, which is defined as ratio of the peak signal to the mean square error of the noise (i.e. noise standard deviation). We systematically tested the dependence of these metrics on the in-plane resolution and the area-normalized noise level η of the system (i.e. noise level per unit area for the chosen sampling frequency).

A one dimensional source with a length of 300 µm in Z direction and at a distance of 50 µm perpendicular to the sensor surface was used to characterize the application scenario of recording from a CA1 subarea. In that case, the pixel size should be not much larger than 10 µm in order to maintain the best possible resolution of the reconstructed signal (Fig. 7B) for the range of tested noise levels η. Increasing the pixel size clearly beyond 10 µm will also tend to decrease the pSNR of the reconstructed axial current density (Fig. 7D). Figs. 8A&B show examples of the magnetic field and reconstructed axial current density at different noise levels. A reasonable reconstruction requires a pSNR > 10, which is exceeded at system noise levels lower than η=10 nTµm.

The analysis was repeated for a source with 1 µm distance to the diamond and a length of 2 µm to characterize the required system parameters when recording from a planar pyramidal cell (Figs. 7C&E; Figs. 8C&D). Here, the pixel size should not be larger than 2 µm in order to maintain the best possible resolution of the reconstructed signal (Fig. 7C). Increasing the pixel size beyond 2 µm will also decrease the pSNR of the reconstructed axial current density (Fig. 7E). In general, system noise levels <~0.4 nTµm are required to achieve a pSNR > 10.

**Discussion**

We have characterized the spatiotemporal characteristics of neural magnetic fields of brain slices through comprehensive computational simulations and theoretical analyses. Our results suggest that peak magnetic field strengths of 2-3 nT can be reached in an experiment, and that the temporal shape is preserved when sampling at a frequency of 800 Hz or higher. The achievable spatial resolution is ~100 µm for $\eta=10$ nTµm, which is the highest noise level of the imaging system that still allows for acceptable axial current density reconstructions, and it improves with decreasing system noise levels. An in-plane resolution of the system of around 10 µm should be chosen to achieve the best resolution and SNR of the reconstructed axial current density distributions.

Our simulations rely on some uncertain parameters and simplifying assumptions that might affect in particular the estimated peak magnetic fields. However, we are confident that their accuracy is good enough in order not to hamper their value to guide further methods development. For example, our simulations assume that all pyramidal cells in the CA1 subarea are recruited by the stimulus, which might suggest higher peak field strengths than occurring in reality. On the other hand, we assume a cell density of 100 cells in a 50 x 50 x 50 µm$^3$ volume, which is a conservatively chosen lower limit, based on the sparse data that is available for the cell density[24,25]. While this indicates that our results might underestimate the peak field strengths, it should be noted that they can be simply linearly rescaled if required. Also our assumption of a 50 µm thick layer of fully dysfunctional cells closest to the diamond sensor is a conservative choice that results in an underestimation of the peak field strengths. Specifically, electrophysiological recordings with single electrodes show that intact pyramidal cell activity can be found already within this layer. Finally, the synchrony of the synaptic events caused by the electrical stimulation was also conservatively estimated, again tending to decrease the peak magnetic fields.

In contrast to imaging the magnetic fields of a brain slice, our results show that imaging of a planar neuron placed on the sensor surface poses more demanding requirements on the system (Figs. 7C&E). This is caused by both lower peak field strengths and by the spatially more confined magnetic fields of the single neuron. The latter property allows for imaging at very high spatial resolutions below 10 µm, but only when very low system noise levels ($\eta < 0.4$ nTµm) can be achieved. In contrast to the reconstruction from the spatially smoother neural magnetic fields expected for a brain slice, noise suppression by the integration of information across a spatial neighborhood, as effectively occurring during Wiener deconvolution, is far less feasible in case of a single planar neural cell.

When diamond sensors are used for sensing low-frequency (DC to kHz) magnetic fields, their sensitivity is fundamentally limited by the NV ensemble density, $n_{NV}$, the dephasing time, $T_2^*$, and the on-resonance fluorescence contrast, $C$ [1,3]. The volume and bandwidth normalized sensitivity limit can be expressed as:

$$\eta_V \cong \frac{h}{g\mu_B} \frac{1}{C\sqrt{\epsilon\, n_{NV}\, T_2^*}},$$

where $\frac{g\mu_B}{h} \cong 28$ Hz/nT is the NV gyromagnetic ratio, and $\epsilon$ is the photon collection efficiency. While this expression is valid for pulsed measurements, continuous wave (CW) techniques typically show a reduced sensitivity due to power broadening. CW techniques, however, are easier to implement in a wide-field imaging setup due to reduced technical requirements. Recently, a volume normalized sensitivity of $\eta_V = 34$ nTµm$^{\frac{3}{2}}$Hz$^{-\frac{1}{2}}$ has been demonstrated using a CW protocol and isotopically engineered diamond with $n_{NV} \sim 1$ ppm and $T_2^* \sim 0.5$ µs[7]. At this level of sensitivity, imaging of the hippocampus tissue with a 5 µm-thick sensing layer and a sampling rate of 1 kS/s corresponds to an area-normalized sensitivity of around 480 nTµm (see Methods eq. (8)). Imaging thus requires averaging of around 2400 trials in order to achieve an area-normalized sensitivity of 10 nTµm of the averaged signal. Further improvement is anticipated from advances in diamond preparation techniques that would lead to a much longer dephasing time, $T_2^* \sim 30$ µs, at a similar NV concentration[26,27]. In order to benefit from such slow-dephasing samples, power broadening has to be avoided and CW protocols can no longer be used. An overall sensitivity improvement of more than two orders of magnitude is expected from combining improved diamond samples with a Ramsey-type measurement scheme[7]. This would allow for planar cell imaging with averages of around $10^2$-$10^3$ trials, while neural activity in brain slices would already be detectable in a single shot. While also the integration of the NV diamond sensor into a functional setup poses additional challenges, our simulation results indicate that the approach suggested here has the potential to reveal important insights into the neural network dynamics in brain slices.

## Methods

*Forward Modeling Scheme for Calculation of the Neural Magnetic Field and Electric Potential*

The simulations of the extracellular neural magnetic fields and electric potential proceeds in two steps. First, the membrane potentials $V_m^n$ and transmembrane currents $I_m^n$ of the simulated neurons are calculated using the NEURON software package (v7.4[12]). For that, NEURON solves the cable equation for complex neural structures, which are discretized into multiple compartments with nonlinear ion channels, and the discretized intracellular and extracellular regions are represented by axial resistances connected by membrane networks [28]. This is schematically depicted in the core-conductor model in Fig. 1C. Then, the extracellular magnetic fields, **B**, and the electric potential, $\Phi$, are determined from the membrane potentials, $V_m^n$, and transmembrane currents, $I_m^n$, using the forward modelling scheme:

$$\mathbf{B}(\mathbf{r},t) = \frac{\mu_0}{4\pi} \sum_{n=1}^{N} \frac{I_{axial}^n(t) \times \hat{\rho}_n}{\rho_n} \left[ \frac{h_n}{\sqrt{h_n^2 + \rho_n^2}} - \frac{l_n}{\sqrt{l_n^2 + \rho_n^2}} \right] \quad (1)$$

$$\phi(\mathbf{r},t) = \frac{1}{4\pi\sigma} \sum_{n=1}^{N} \frac{I_m^n(t)}{\Delta s_n} \log \left| \frac{\sqrt{h_n^2 + \rho_n^2} + h_n}{\sqrt{l_n^2 + \rho_n^2} + l_n} \right| \quad (2)$$

with $I_{axial}^n = \frac{V_m^n - V_m^{n-1}}{\Delta s_n r_i^n}$. It is assumed that the neuron is divided into N compartments and $\Delta s_n$ denotes the length of each compartment. In addition, $\rho_n$ represents the distance perpendicular to the orientation of the main axis of the compartment, $h_n$ the longitudinal distance from the end of the compartment, and $l_n = h_n - \Delta s_n$ the longitudinal distance from the start of the compartment. The extracellular potentials are determined using the LFPy toolbox[29] and the magnetic field is calculated by adapting LFPy to evaluate eq. (2). Further details on the forward modeling scheme are summarized in the Supplementary Material, which also covers the validation of our simulation approach by comparing it with previously presented in vitro results [7]. The employed forward scheme was additionally validated by comparison with the analytical solution for the case of a long straight axon, which is also summarized in the Supplementary Material.

The total magnetic field and electric potential created by activation of the CA1 subarea are determined by calculating the field for each contained pyramidal cell, and the final result is obtained by a linear superposition of the results of the single cells. For that, the 500 µm x 500 µm x 300 µm large CA1 subarea is divided into several layers with 50 µm thickness along the Z direction, and 1000 pyramidal cell models are placed in each layer [24,25], with the main dendrites

11/39

pointing in the Y direction. The soma locations are randomly jittered within the layer, with the jitter being restricted to a 50 µm wide band in Y direction. As the same cell model was used for all locations, the cells are rotated randomly around the orientation of their main dendrite in order to prevent the formation of artefacts created by the field superposition of the single cells. In the literature, the neural density of pyramidal cells is given either as total number ($4 \times 10^5$ cells in CA1, corresponding to $2.25 \times 10^6$ cells/mm$^3$) [24] or as density in the pyramidal cell layer ($1.74 \times 10^6$ cells/mm$^3$) [25]. These numbers suggest 281 cells [25] and 217 cells [24] in a 50 µm x 50 µm x 50 µm volume, based on the assumption that the soma of the pyramidal cells are contained in a 50 µm wide band in Y direction, denoted as the "pyramidal layer". The experimental studies did not distinguish different neural cell types in the pyramidal layer when determining the cell counts, and while pyramidal cells are the most frequent cell type in that layer, using the above results without correction would overestimate the number of pyramidal cells. We therefore base our simulations on 100 cells in a 50 µm x 50 µm x 50 µm volume, which is a conservative lower limit. However, the neuronal density varies for different slices and species, and if desired, our results can be easily linearly rescaled to match those for a different neuron density. We consider only pyramidal cells for our simulations, as their regular and longitudinal cell morphologies result in well detectable external fields which sum up across neighboring cells. The external magnetic fields and electric potential of layered cortical structures are thus dominated by the responses of this neural cell type [20].

*Achievable spatial resolution and required in-plane resolution and sensitivity of the sensor*

The reconstruction of the neural axial current density (representing the current flow within the neurons) in the brain slice from the measured 2D magnetic field distributions allows us to determine the spatial position and extent of the activated hippocampal subarea. The current density reconstruction has the form of a 2D deconvolution problem, for which Wiener deconvolution is the optimal method in case of additive noise[23]. It applies an inverse Wiener filter in the spatial frequency domain to "undo" the effects of the magnetic field that acts as a spatial low-pass filter of the neural axial current density (Fig. 7A). We are interested to determine the maximally achievable spatial resolution and the signal-to-noise ratio of the reconstructed current density. In particular, in order to guide the development of the measurement setup, we aim to characterize their dependence on the in-plane resolution of the system and on its signal-to-noise ratio.

In the following, we consider the magnetic field of a 2D point source and determine the full width at half maximum (FWHM) and the peak-signal-to-noise ratio (pSNR) of the current density reconstructed by inverse Wiener filtering. Given the orientation of the pyramidal cells along the Y direction in the modelled CA1 subarea (Fig. 1B), the $J_X$ and $J_Z$ components of the neural axial current densities are small and their contribution to the magnetic field can be neglected. The neural source currents $J_Y$ result in a magnetic field distribution at the diamond surface that has only $B_X$ and $B_Z$ components. We focus on the $B_X$ component of the field, as it has much higher peak strengths (Fig. 4). In addition, we assume a homogenous excitation of the pyramidal cells along the full height of the CA1 subarea (except for the layer of dead cells), so that the strength of the neural axial current density $J_Y$ is constant across the depth of the tissue. Denoting the thickness of the layer of dead cells by $z_0$ and the thickness of the layer of active cells as d, the $B_X$ component of the magnetic field at the diamond surfaces is:

$$B_X(x,y) = \frac{\mu_0}{4\pi} \int_0^d \int_{-\infty}^{\infty} \int_{-\infty}^{\infty} \frac{J_y(x',y')(z_0 + z')}{[(x-x')^2 + (y-y')^2 + (z_0+z')^2]^{\frac{3}{2}}} dx' dy' dz' . \tag{3}$$

Transformation to the spatial frequency domain and integration along Z yields the following simplified relation[30]:

$$b_x(k_x, k_y) = \mu_0 \exp\left(-\left(z_0 + \frac{d}{2}\right)\sqrt{(k_x^2 + k_y^2)}\right) \frac{\sinh\left(\frac{d}{2}\sqrt{(k_x^2 + k_y^2)}\right)}{\sqrt{(k_x^2 + k_y^2)}} j_y(k_x, k_y) \tag{4}$$

$$= f(k_x, k_y) j_y(k_x, k_y) ,$$

where $k_x$ and $k_y$ are spatial frequencies. Equation (4) shows that the magnetic field acts as a spatial low-pass filter of the axial current density. Its cut-off frequency decreases with increasing distance to the sensor and increasing slice thickness. Considering that the magnetic field stated in eq. (4) is the convolution function of the system (or source-to-measurement transformation filter), the Wiener deconvolution filter is given by [23]

$$f^I(k_x, k_y) = \frac{\overline{f}(k_x, k_y)}{|f(k_x, k_y)|^2 + \frac{s_\eta(k_x, k_y)}{s_j(k_x, k_y)}}, \qquad (5)$$

with the overbar indicating the complex conjugate, $s_j$ denoting the power spectrum density (PSD) of the axial current density and $s_\eta$ the PSD of the noise. The ratio between $s_\eta$ and $s_j$ controls the amount of regularization that is applied to the deconvolution.

As a next step, we assume a point source with known signal power, and relate the ratio between $s_\eta$ and $s_j$ to this signal power and to known system parameters. Assuming spatially uncorrelated white Gaussian noise, $s_\eta$ is constant across spatial frequencies. The noise power of the image is then given as

$$\sigma_n^2 = \frac{1}{(2\pi)^2} \int_{-k_{sy}}^{k_{sy}} \int_{-k_{sx}}^{k_{sx}} s_\eta \, dk_x dk_y = \frac{s_\eta k_{sx} k_{sy}}{\pi^2}, \qquad (6)$$

where $k_{sx}, k_{sy}$ are the maximum spatial frequencies in dependence of the in-plane resolution of the image ($k_{sx} = \pi/\Delta_x$; $k_{sy} = \pi/\Delta_y$; $\Delta_x = \Delta_y = \Delta$).

The noise power of the image can also be written in terms of the pixel-normalized noise level $\eta_{pixel}$, the number of pixels MxM (for a 2D grid with M pixels in each direction) and the pixel area $\Delta^2$ as

$$\sigma_n^2 = \sum_{j=1}^{M} \sum_{i=1}^{M} \eta_{pixel}^2 \Delta^2 = \eta_{pixel}^2 \Delta^2 M^2, \qquad (7)$$

with

$$\eta_{pixel} = \frac{\eta_V}{\sqrt{V_{pixel}}} \sqrt{f_s} = \frac{\eta_V}{\Delta \sqrt{h}} \sqrt{f_s} = \frac{\eta}{\Delta}, \qquad (8)$$

where $f_s$ is the sampling frequency, and $V_{pixel}$ the volume and $h$ the height of the NV layer, respectively. Equation (8) relates the pixel-normalized noise level $\eta_{pixel}$ to the volume-normalized sensitivity $\eta_V$, which is usually reported to characterize the sensitivity of the system (see e.g. [7]). It further relates it to the area-normalized noise level $\eta$ (i.e. noise level per unit area for the chosen sampling frequency), which we define to account for the fact that the height of the NV layer and the sampling frequency were already selected and are kept constant. Both $\eta_V$ and $\eta$ are independent of the chosen in-plane resolution, and they can be determined via measurements.

Equations (6 - 8) can then be used to determine $s_\eta$ from the area-normalized noise level $\eta$ of the system:

$$s_\eta = \eta^2 M^2 \Delta^2 = \eta^2 A_{FoV}. \tag{9}$$

If $j_Y(x,y)$ is assumed to be a single dipole source having a dipole moment $\sigma_j$, its source power spectral density will be:

$$s_j(k_x, k_y) = |FT\{j_y(x,y)\}|^2 = \sigma_j^2. \tag{10}$$

Incorporating these results into the Wiener deconvolution filter (eq. (5)) yields:

$$f^I(k_x, k_y) = \frac{\bar{f}(k_x, k_y)}{|f(k_x, k_y)|^2 + \frac{\eta^2 A_{FoV}}{\sigma_j^2}}, \tag{11}$$

which indicates that the filter expression is independent of the chosen pixel size. Applying the Wiener filter to reconstruct the current density of the point source for a given pixel size $\Delta$ finally gives:

$$\hat{j}(x,y) = \frac{1}{(2\pi)^2} \int_{-k_{sy}}^{k_{sy}} \int_{-k_{sx}}^{k_{sx}} \sigma_j\, f^I(k_x, k_y) f(k_x, k_y)\, e^{i(k_x x + k_y y)} dk_x dk_y$$

$$=$$

$$\frac{1}{(2\pi)^2} \int_{-\infty}^{\infty} \int_{-\infty}^{\infty} \left(\sigma_j\, f^I(k_x, k_y) f(k_x, k_y)\right) \left(\mathrm{rect}\left(\frac{k_x}{2k_{sx}}\right) \mathrm{rect}\left(\frac{k_y}{2k_{sy}}\right)\right) e^{i(k_x x + k_y y)}\, dk_x dk_y \tag{12}$$

$$= \sigma_j\, FT^{-1}\left\{\frac{|f(k_x, k_y)|^2}{|f(k_x, k_y)|^2 + \frac{\eta^2 A_{FoV}}{\sigma_j^2}}\right\} * \frac{1}{\Delta^2}\mathrm{sinc}\left(\frac{x}{\Delta}\right)\mathrm{sinc}\left(\frac{y}{\Delta}\right)$$

where $*$ represents the convolution operator. The reconstructed current density of the point source is equivalent to the point spread function (PSF) of the system, and the maximally achievable spatial resolution can be characterized using the FWHM of the PSF. The first term on the right side depends on parameters of the point source (its strength, height and distance to the sensor), the area-normalized noise level and the area of the field of view. The second term depends only on the discretization parameter, i.e. the chosen in-plane resolution. Therefore, their impact on the system resolution can be analyzed separately. For example, as the pixel size decreases, the second term approaches a delta distribution and the resolution will be determined by the first term.

The first term is a low pass filter which increases in smoothness with increasing noise level. On the other hand, for large pixels, the second term dominates and determines the resolution limit of the system.

## Acknowledgements

This work received financial support through the EXMAD project funded by the Innovation Fund Denmark and the Novo Nordisk Foundation.

## Author contributions

MK performed the neural simulations and analysis of the results. MK and AT developed the modeling and analysis framework. NOD and AMW supported the design of the computational experiments and the data interpretation. All authors contributed to writing the paper. The project was conceived by AT, AH and ULA.

## Additional information

**Competing financial interests:** The authors declare no competing financial interests.

# Figure legends

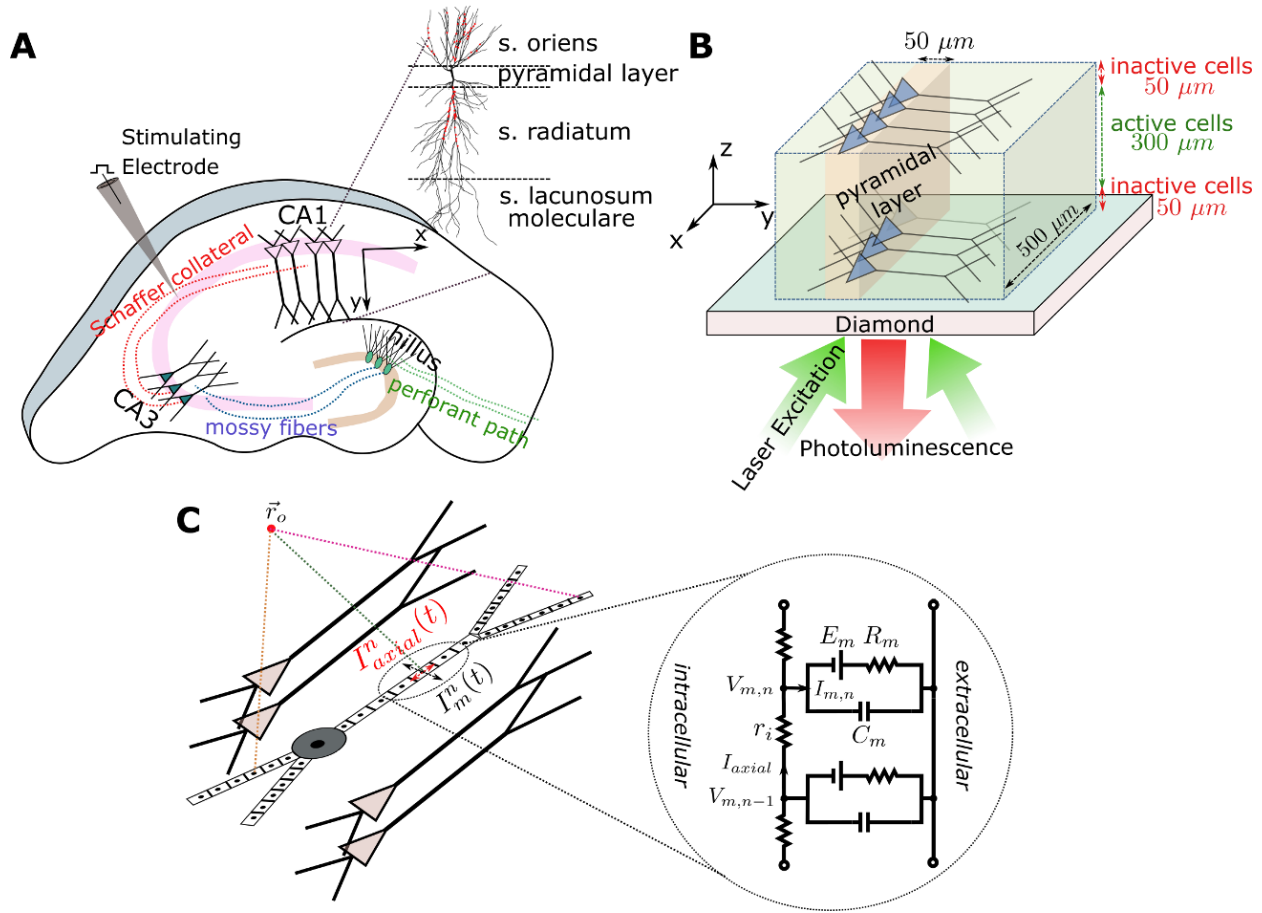

**Figure 1:** **(A)** Illustration of a hippocampus slice with its trisynaptic path. We consider the recording of neural activity from the CA1 area that is evoked by the electrical stimulation of the Schaffer collaterals. **(B)** Schematic illustration of the simulated CA1 subarea with a size of 500 x 500 x 300 µm³ placed on the diamond sample. It is assumed that the neural cells in a distance of up to 50 µm to the diamond are dysfunctional due to the preparation. The pyramidal cells are equally distributed in the patch along the X and Z directions, and their soma locations are randomly jittered in a 50 µm wide band in Y direction. **(C)** Multi-compartment model of a pyramidal cell, as applied in our forward modeling scheme.

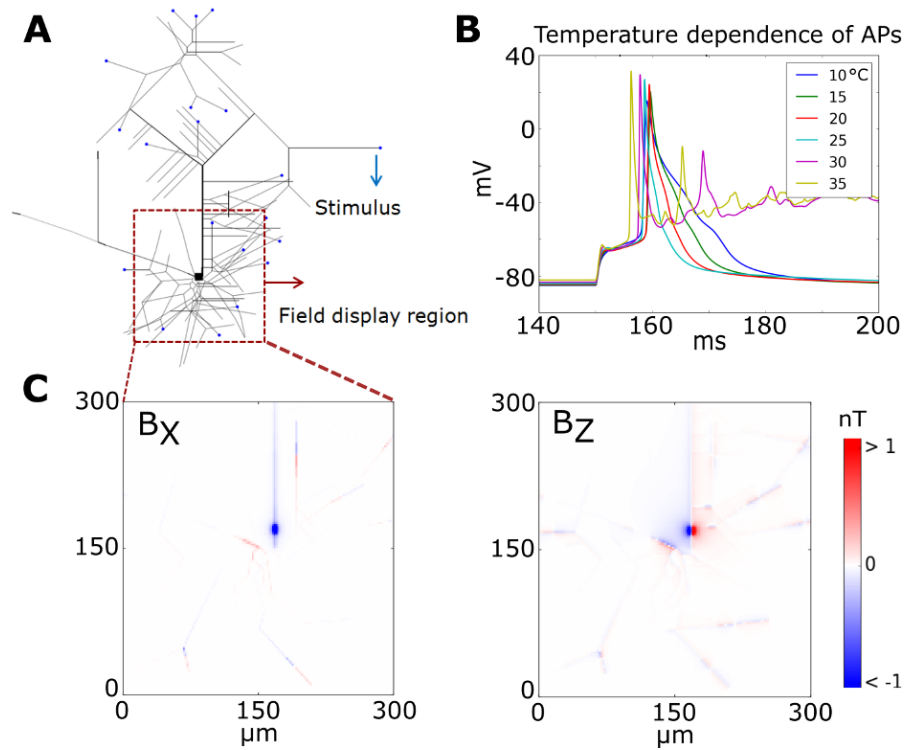

**Figure 2:** **(A)** Morphology of a single planar CA1 pyramidal cell and the current injection sites (blue dots). The cell model has 265 anatomical sections and 15 different voltage active channels. **(B)** Transmembrane potentials generated at the soma as a function of time for different temperatures. The duration of the action potential increases as the temperature decreases. **(C)** Zoomed plots of the peak $B_X$ and $B_Z$ components of the magnetic field close to the soma. The $B_Y$ component is negligibly small. The simulated fields are assessed at a distance of 100 nm to the surface of each branch and determined at an ambient temperature of 25°C.

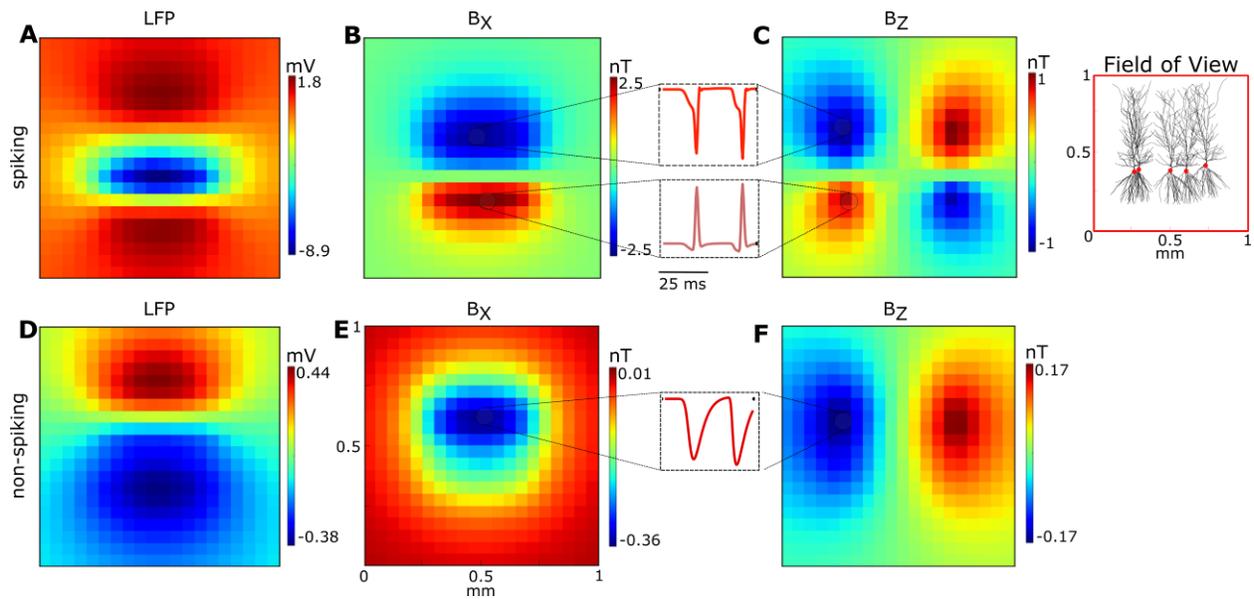

**Figure 3:** Simulation results showing the spatial distribution and the temporal shape of the extracellular fields for the CA1 subarea for the **spiking case** (top row, A-C) and the **non-spiking case** (bottom row, D-F). The inset on the right exemplarily depicts the size and orientation of the pyramidal cells in relation to the imaged field of view. The spatial distributions are extracted at the time point when the field peaks. (**A**) The LFP for the **spiking** case has a large negative peak close to the soma layer, as the generation of the action potentials causes a strong current inflow. (**B**) $B_X$ component of the magnetic field (spiking case). The generation of the action potentials creates separate strong axial current flows in opposite directions from the S.O. and S.R. dendrites to the soma of the pyramidal cells, which leads to a negative peak in S.R. and positive peak in S.O. regions. (**C**) $B_Z$ component of the magnetic field (spiking case), in accordance to the two opposite axial current flows from the S.O. and S.R. dendrites to the soma regions. The **insets** depict the temporal shapes of the magnetic field components, extracted from the indicated positions (a time window of 50 ms is shown). The initial phases before the action potentials reflect the accumulation of excitatory postsynaptic potentials (EPSPs). (**D**) The LFP for the **non-spiking** case has a dipolar distribution, with the negative peak caused by the EPSPs in the S.O. region and the positive peak above S.R. region caused by outward membrane currents which balance the excitatory synaptic currents. (**E**) $B_X$ component of the magnetic field (non-spiking case). It has a large peak around the soma, indicating an axial current flow from the S.R. to the S.O. region. (**F**) $B_Z$ component of the magnetic field (non-spiking case), in accordance to the axial current flow from S.R. to S.O. The temporal shapes depicted in the insets reflects the accumulation of EPSPs.

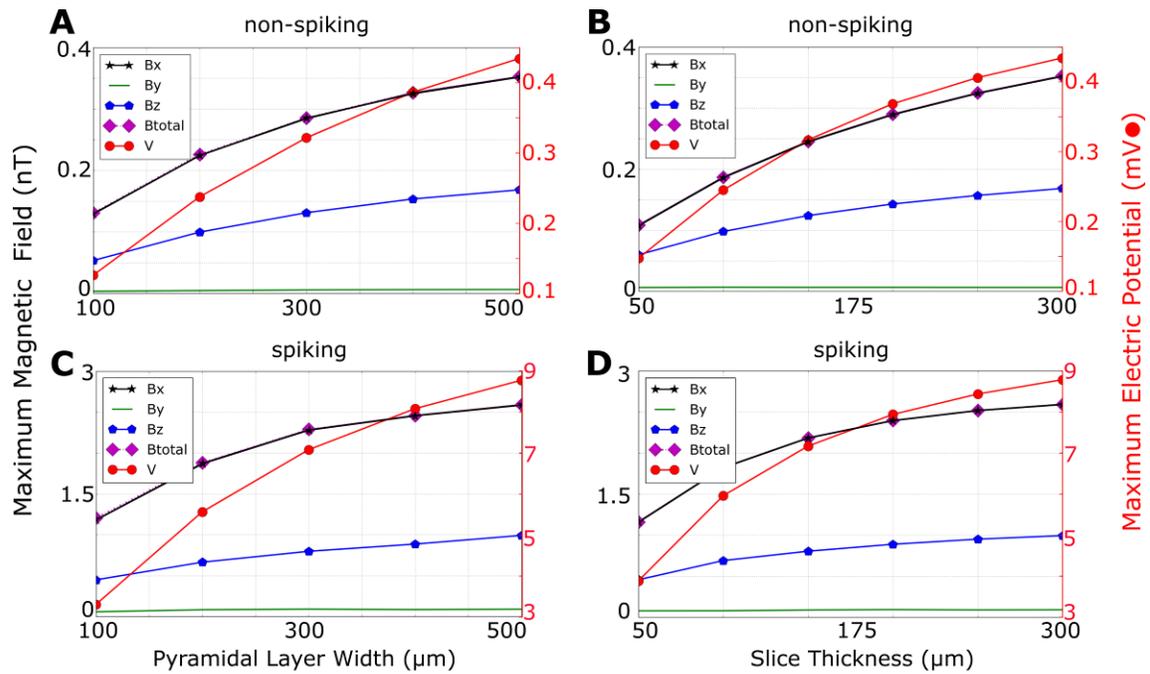

**Figure 4:** Dependence of the maximum field strength on the size of the active CA1 subarea. The distance of the active subarea to the sensor surface is 50 µm in all simulations. (**A**) Non-spiking case. The width of the subarea (X direction in Fig. 1B) is varied from 100 µm to 500 µm, while the thickness is held constant at 300 µm. (**B**) Non-spiking case. The thickness of the subarea (Z direction in Fig. 1B) is systematically varied from 50 µm to 300 µm. The width is held constant at 500 µm. (**C**) Spiking case. The width of the subarea is varied from 100 µm to 500 µm, the thickness is 300 µm. (**D**) Spiking case. The thickness of the subarea is systematically varied from 50 µm to 300 µm, the width is 500 µm.

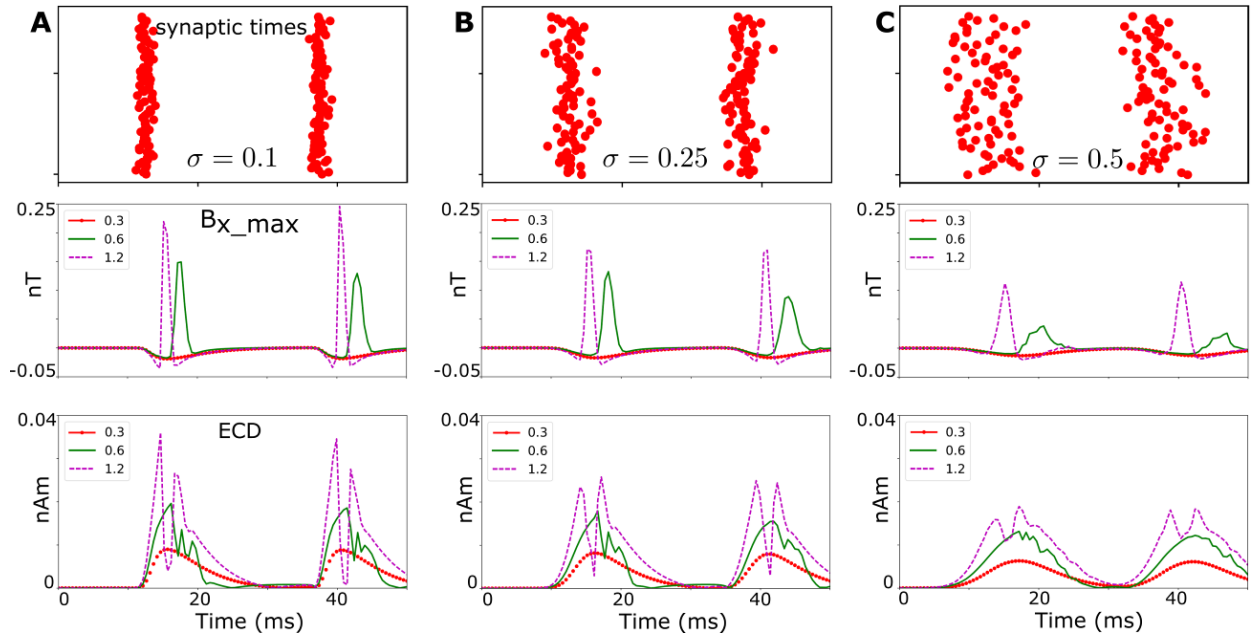

**Figure 5:** Dependence of the field strength on the temporal synchrony of the EPSPs and on the simulated strength of the glutamatergic synapses. A sublayer is simulated that contains 200 cells and has a width of 100 µm, a height of 50 µm, and a minimal distance of 150 µm to the diamond surface. The time points of the synaptic excitations are jittered in time using a Gaussian profile, determined by standard deviation σ. The synaptic strengths were varied between 0.3 nS (resulting in non-spiking activity), 0.6 nS (just resulting in stable spiking activity) and 1.2 nS (resulting in more synchronous spiking). The top row depicts the temporal distribution of the synaptic excitations. The middle row shows the time course of the $B_X$ component of the magnetic field, extracted at the spatial peak position. The bottom row shows the time course of the equivalent current dipole (ECD), again extracted at the spatial peak position. (**A**) Results for highly synchronous synaptic excitations ($\sigma = 0.1$). (**B**) Synchronous synaptic excitations ($\sigma = 0.25$). (**C**) Weakly synchronous excitations ($\sigma = 0.5$).

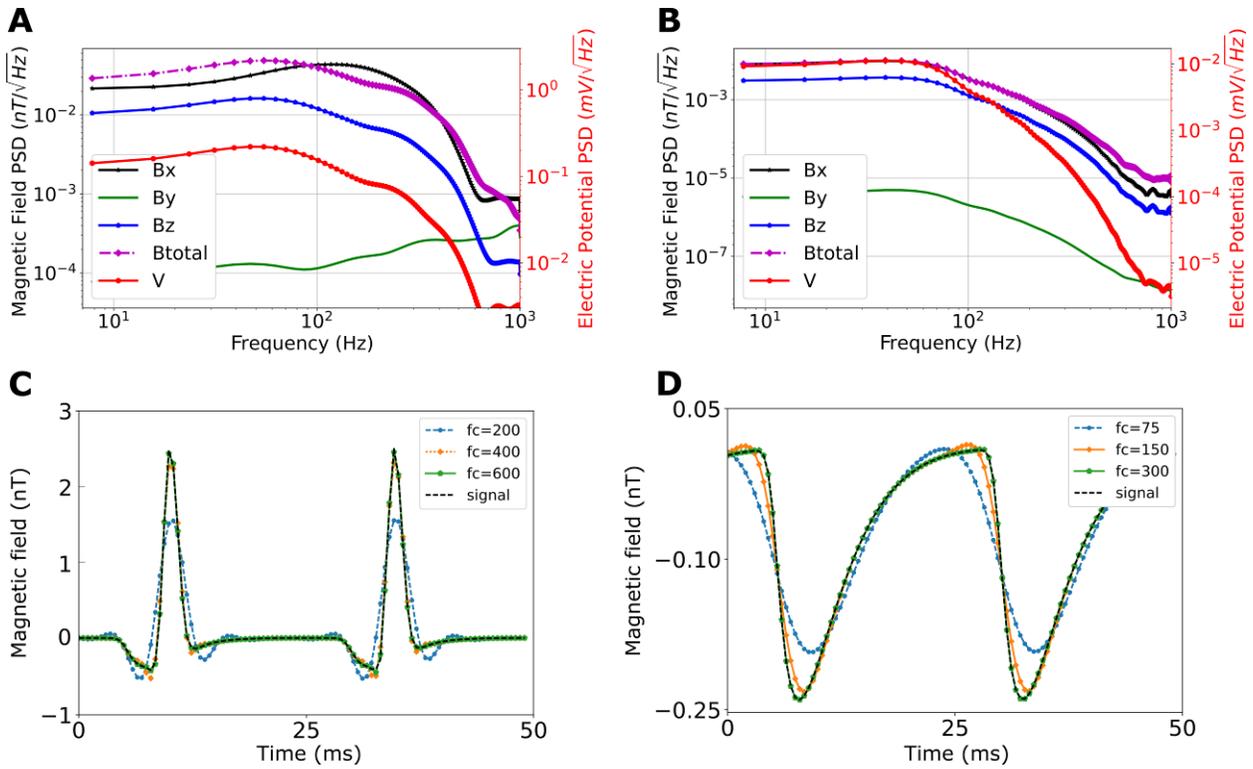

**Figure 6:** **(A)** Power spectral densities of the magnetic field components, the total magnetic field and the electric potential for the **spiking** case of the 500 x 500 x 300 µm³ large CA1 subarea (50 µm distance to the sensor). **(B)** Power spectral densities for the **non-spiking** case. The power spectral density of the LFP is shown for comparison. **(C)** Corresponding temporal waveforms for the spiking case, both non-filtered and low-pass filtered with a third-order Butterworth filter at different cut-off frequencies. The signal shape is well preserved for cut-off frequencies of 400 Hz and higher. **(D)** Temporal waveforms for the non-spiking case. The signal is well preserved for cut-off frequencies of 150 Hz and higher. The time distributions in **C** and **D** are extracted at the spatial point where the maximum field occurs.

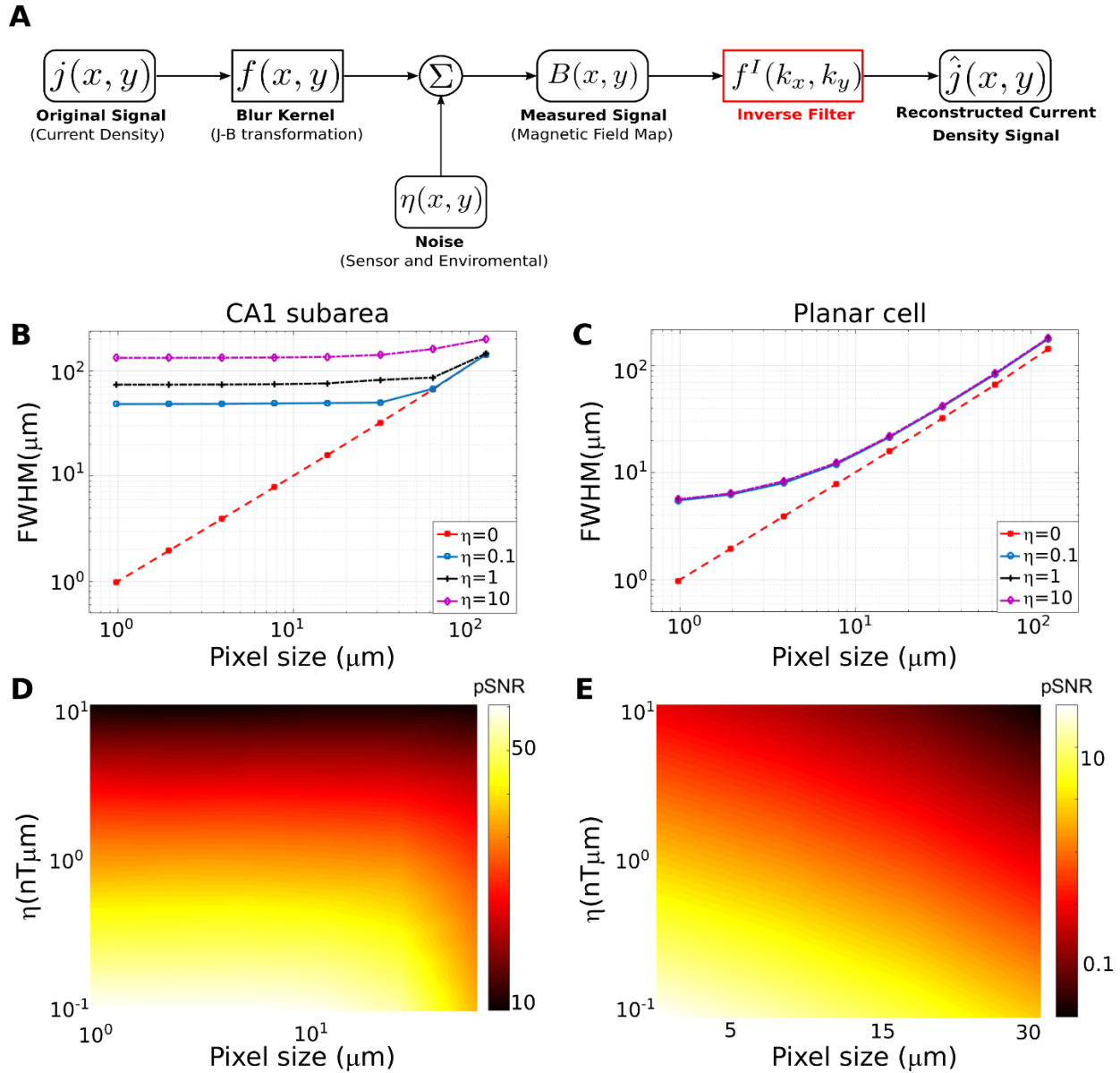

**Figure 7:** **(A)** Schematic outline of the reconstruction of neural axial current densities from the measured magnetic fields. **(B)** Recording from the CA1 subarea. Shown is the achievable system resolution (characterized by the full width at half maximum – FWHM – of the reconstructed neural axial current density) in dependence of the in-plane resolution of the sensor. The neural source is 300 µm long at a distance of 50 µm from the diamond surface and pointing in the Z direction perpendicular to the sensor plane. The source strength was selected so that the magnetic peak field strength was 2.5 nT at the diamond surface, thereby matching the peak field strength observed in the simulations of the CA1 subarea. The achieved system resolution is almost constant for higher

in-plane resolutions, but drops off quickly for too low in-plane resolutions. **(C)** Recording of a planar cell. Shown is the achievable system resolution in dependence on the in-plane resolution. The distance of the neural point source from the diamond surface is 1 µm and the length is 2 µm. The source strength was again selected to yield a peak magnetic field of 2.5 nT at the diamond surface, in order to allow for a direct comparison with the results obtained for the CA1 subarea. Applying the results to the situation of a single planar pyramidal cell thus requires rescaling, as our neural simulations indicate peak magnetic fields of 1 nT for the latter. Except for very high in-plane resolutions of $< 4$ µm, the system resolution depends linearly on the in-plane resolution. **(D)** Peak-signal-to-noise ratio (pSNR) for recordings from the CA1 subarea as a function of the in-plane resolution of the sensor (pixel size) and the signal-to-noise power of the source, $\eta$. As long as the in-plane resolution is chosen high enough to maintain the best possible system resolution, also the pSNR of the reconstructed axial current densities maintain constant. **(E)** Peak-signal-to-noise ratio (pSNR) for recordings from a planar cell.

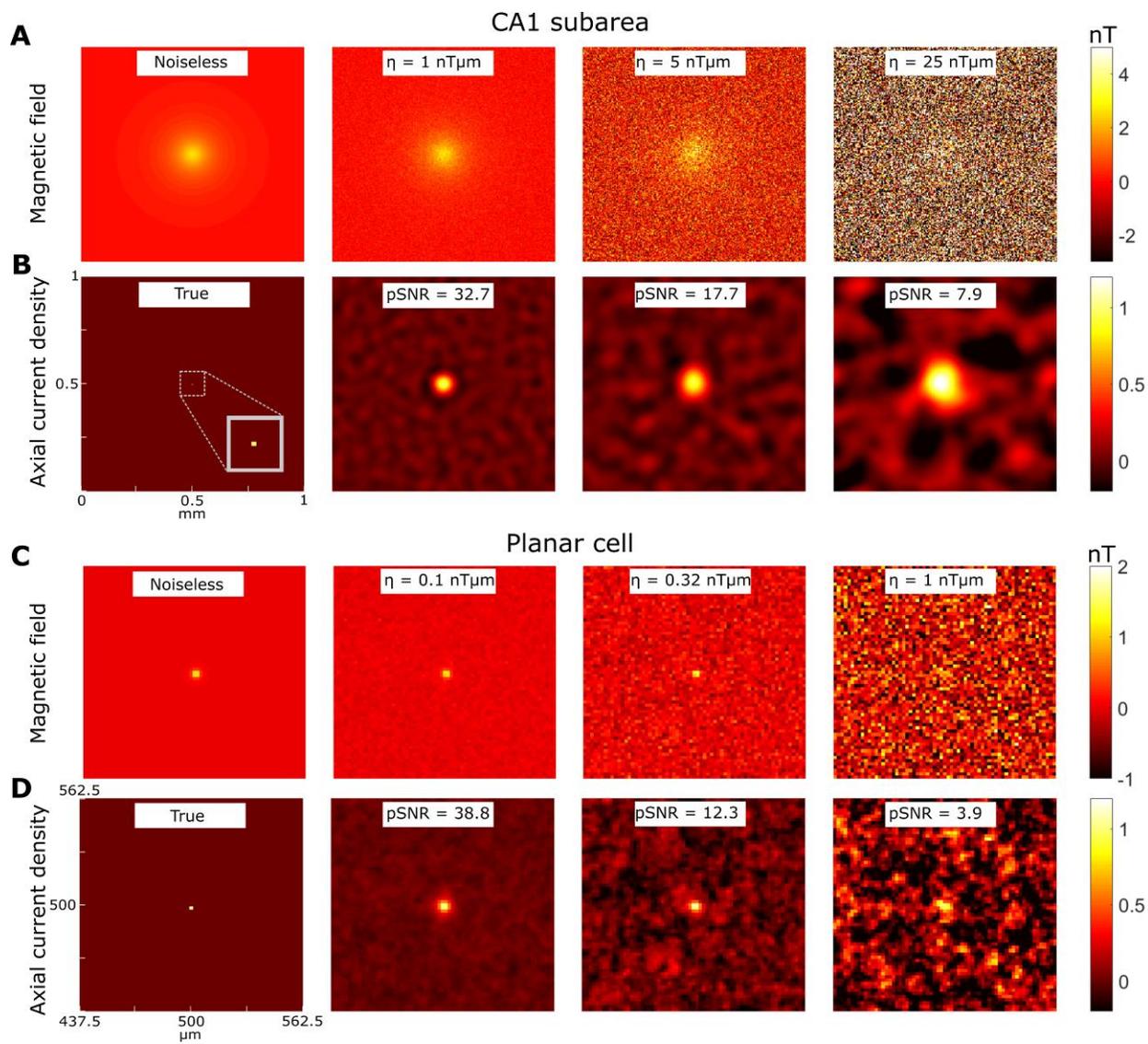

**Figure 8:** Examples of reconstructed neural axial current densities at varying pSNR levels. **(A)** Recording from the CA1 subarea. A neural point source with 50 µm distance to the diamond plane and a length of 300 µm in Z direction is simulated. The strength of the source is chosen so that a peak magnetic field strength of 2.5 nT is reached in the sensor plane. The in-plane resolution of the sensor is set to 7.8 µm (128x128 pixels in 1mm$^2$). The $B_X$ component of the measured magnetic field is shown for the noiseless case and for three different levels of the signal-to-noise power of the recorded magnetic field. **(B)** The normalized axial current densities reconstructed from the corresponding magnetic fields shown in A. Each plot is normalized to the maximum of its noiseless reconstruction with the same spatial filter settings. **(C)** Recording of a planar cell. A neural point source with 1 µm distance to the diamond and a length of 2 µm is simulated. Its strength is chosen

to give a peak magnetic field strength of 2.5 nT in the sensor plane. The $B_X$ component of the measured magnetic field is shown for the noiseless case and for three different levels of the signal-to-noise power. The in-plane resolution of the sensor is set to 2 µm (512x512 pixels in 1mm$^2$) and a field of view of 125x125 µm$^2$ is selected for better visualization. Since the magnetic field is very localized, the "recorded" peak magnetic field strength is slightly lower than 2.5 nT due to spatial averaging in the 2 µm x 2 µm areas of the sensor pixels. **(D)** The normalized axial current densities reconstructed from the corresponding magnetic fields in C. Each plot is normalized to the maximum of its noiseless reconstruction with the same spatial filter settings.

**M Karadas, A M Wojciechowski, A Huck, N O Dalby, U Lund Andersen, A Thielscher**

**Opto-magnetic imaging of neural network activity in brain slices at high resolution using color centers**

**Supplementary Material**

**Forward Modeling Scheme**

In the following, we describe the employed forward modelling scheme to determine the extracellular electric potential from the transmembrane currents and to determine the extracellular magnetic field from the intracellular axial currents, i.e. the current flow parallel to the cell membrane. For a cylindrical nerve structure embedded in a large extracellular medium (Fig. S1), the extracellular volume $V_e$ can be approximated as infinite homogeneous conductor, so that the cell membrane is the only boundary, denoted by *S*. In that case, the electric potential in the extracellular region can be written as (starting from eq. 26 and 27 given in [1])

$$\phi_e(\mathbf{r}) = \frac{1}{4\pi\sigma_e} \int_{V_e} \frac{\nabla' \mathbf{J}_{\text{source}}(\mathbf{r}')}{|\mathbf{r} - \mathbf{r}'|} dv' \quad \text{(S1)}$$

$$+ \frac{1}{4\pi\sigma_e} \int_{S} (\sigma_i \phi_i - \sigma_e \phi_e) \nabla' \frac{1}{|\mathbf{r} - \mathbf{r}'|} \mathbf{n} \, dS'$$

where operator $\nabla'$ acts on $\mathbf{r}'$. Since there are no current sources in the extracellular medium (i.e., $\nabla \mathbf{J}_{\text{source}} = 0$ in the extracellular space), the first term on the right hand side disappears. The extracellular potential is far smaller than the intracellular potential [2] (i.e., $\sigma_i \phi_i \gg \sigma_e \phi_e$), so that the above equation can be approximated as:

$$\phi_e(\mathbf{r}) = \frac{1}{4\pi\sigma_e} \int_{S} \sigma_i \phi_i \nabla' \frac{1}{|\mathbf{r} - \mathbf{r}'|} \mathbf{n} \, dS' \quad \text{(S2)}$$

Therefore, the extracellular potential is approximately independent of the external current distribution. If the nerve structure has a small cross-section A, so that we can assume that $\phi_i$ is only changing along the structure (this direction is parameterized using the longitudinal coordinate l), the surface integral can be written in terms of a volume integral[2]:

$$\phi_e(\mathbf{r}) = \frac{1}{4\pi\sigma_e} \int_A \left[ \int_{-\infty}^{\infty} \sigma_i \frac{\partial^2 \phi_i}{\partial l^2} \frac{1}{|\mathbf{r}-\mathbf{r'}|} dl \right] dA' \tag{S3}$$

The numerical simulation of the neuron are based on core-conductor equations, which assume a constant current density across the cross-section of the cell, so that

$$\frac{\partial^2 \phi_i}{\partial l^2} = r_i i_m \tag{S4}$$

where $r_i = 1/\sigma_i A$ is the intracellular resistance per unit length and $i_m$ is the membrane current per unit length. Equation (S4) allows us to write the extracellular potential in dependence on the membrane currents:

$$\phi_e(\mathbf{r}) = \frac{1}{4\pi\sigma_e} \int_{-\infty}^{\infty} i_m \frac{1}{|\mathbf{r}-\mathbf{r'}|} dl \tag{S5}$$

The core-conductor model provides a good approximation as we are only interested in the transmembrane potential of the cell model[3]. Representing the nerve geometry by a series of N cylindrical compartments of length allows discretizing eq. (S5) as

$$\phi(\mathbf{r}, t) = \frac{1}{4\pi\sigma_e} \sum_{n=1}^{N} \int_{L_n} \frac{I_m^n(t) dl}{\Delta s_n |\mathbf{r}-\mathbf{r}_n|} \tag{S6}$$

where $I_m^n$ is the membrane current for compartment n, $\Delta s_n$ is its length, and a is its radius:

$$I_m^n = i_m \Delta s_n = \frac{I_m \Delta s_n}{2\pi a} \tag{S7}$$

Integrating eq. (S6) along the center line of each compartment finally gives

$$\phi(\mathbf{r}, t) = \frac{1}{4\pi\sigma_e} \sum_{n=1}^{N} \frac{I_m^n(t)}{\Delta s_n} \log \left| \frac{\sqrt{h_n^2 + \rho_n^2} + h_n}{\sqrt{l_n^2 + \rho_n^2} + l_n} \right| \tag{S8}$$

where $\rho_n$ is the distance perpendicular to the center line of the compartment, $h_n$ is the longitudinal distance from the end of the compartment, and $l_n = h_n - \Delta s_n$ denotes the longitudinal distance from the start of the compartment [4].

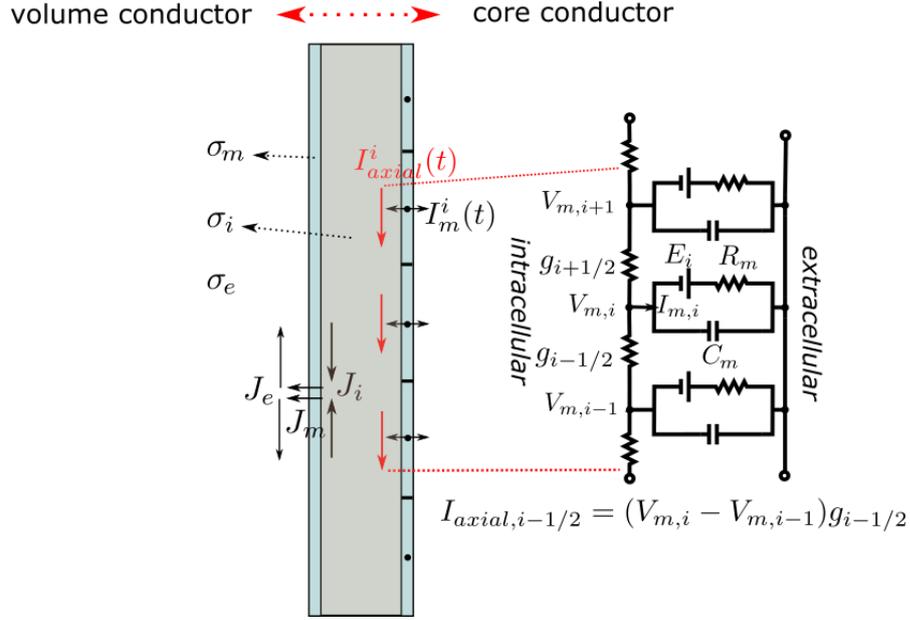

**Figure S1:** Model of a cylindrical neural process. The left side shows the volume-conductor model and the right side the equivalent core-conductor model. The volume conductor model distinguishes between the intracellular, membrane and extracellular regions. In the core-conductor model, the intracellular and extracellular regions are represented by axial resistances and they are connected by a membrane network[5].

A similar solution can be derived for the extracellular magnetic field (starting from eq. 20 and 22 in [1]) :

$$\mathbf{B}(\mathbf{r}) = \frac{\mu_0}{4\pi}\left\{\int_V \frac{\mathbf{J}_{\text{source}}(\mathbf{r}') \times (\mathbf{r}-\mathbf{r}')}{|\mathbf{r}-\mathbf{r}'|^3}\,dv' - \int_V \frac{\sigma(\mathbf{r}')\nabla'\phi(\mathbf{r}') \times (\mathbf{r}-\mathbf{r}')}{|\mathbf{r}-\mathbf{r}'|^3}\,dv'\right\} \qquad (S9)$$

The domain is divided into the intracellular, membrane and extracellular volumes, and the integral equations run over the complete domain. Since the extracellular and intracellular volumes are source free ohmic regions, the first integral is nonzero only at the membrane. We thus define $\mathbf{J}_{\text{source}} = \mathbf{J}_\mathbf{m}$ as the biochemical current sources in the membrane. The contribution of the secondary currents in the second integral of eq. (S9) can be written for the intracellular and extracellular regions by considering vector identities as:

$$\int_{V_{i,e}} \frac{\sigma_{i,e}\nabla'\Phi(\mathbf{r}') \times (\mathbf{r}-\mathbf{r}')}{|\mathbf{r}-\mathbf{r}'|^3}\,dv' \qquad (S10)$$

$$= \int_{V_{i,e}} \frac{\nabla' \times (\sigma_{i,e}\nabla'\Phi(\mathbf{r}')) \times (\mathbf{r}-\mathbf{r}')}{|\mathbf{r}-\mathbf{r}'|^3}\,dv'$$

$$+ \oint_{S_{i,e}} \frac{\sigma_{i,e}\nabla'\Phi(\mathbf{r}') \times \hat{n}}{|\mathbf{r}-\mathbf{r}'|}\,ds'$$

For piecewise homogeneous regions, the volume integral on the right side is zero. Therefore, the second volume integral of eq. (S9) can be converted into surface integrals. The magnetic field can be expressed as sum of volume and surface integrals over the membrane [6]:

$$\mathbf{B}(\mathbf{r}) = \frac{\mu_0}{4\pi} \left\{ \int_{v_m} \frac{\mathbf{J_m}(\mathbf{r}') \times (\mathbf{r} - \mathbf{r}')}{|\mathbf{r} - \mathbf{r}'|^3} dv' \right. \tag{S11}$$

$$\left. + \oint_{S_e} \frac{\sigma_e \nabla' \phi(\mathbf{r}') \times \hat{n}}{|\mathbf{r} - \mathbf{r}'|} ds' + \oint_{S_i} \frac{\sigma_i \nabla' \phi(\mathbf{r}') \times \hat{n}}{|\mathbf{r} - \mathbf{r}'|} ds' \right\}$$

The current densities are symmetric along the nerve structure, so they can be written in terms of their radial components $J^r$ and longitudinal components $J^l$

$$\mathbf{J}(r, l) = J^r(r, l)\hat{r} + J^l(r, l)\hat{l} \tag{S12}$$

with $\hat{r}$ and $\hat{l}$ being unit vectors in radial and axial direction. The membrane conductivity is extremely small with respect to the conductivities of the inner and outer regions, so that we can assume that the membrane currents $\mathbf{J_m}$ have only a radial direction[6]. As the membrane is very thin, the contribution of the membrane source to the magnetic field can be safely neglected and the volume integral in eq. (S11) can be skipped.

At the inner and outer surfaces of the membrane, the radial components of the extracellular and intracellular current densities cannot contribute to the surface integrals, since they are orthogonal to the surface normal $\hat{n}$. Therefore, the total magnetic field is determined by the longitudinal (axial) components of currents $J_e$ and $J_i$ at the outer and inner surfaces, which are defined as:

$$J_{e,i} = \sigma_{e,i} \frac{\partial \Phi_{e,i}}{\partial l} \tag{S13}$$

It can be further assumed that the external current flow $\mathbf{J_e}$ is distributed in a large volume, so that its longitudinal component at the outer membrane surface is small. In contrast, the internal current flow $\mathbf{J_i}$ is confined to a small volume, so that its longitudinal component is strong and dominates eq. (S11). Thus, equation (S11) can be approximated as:

$$\mathbf{B}(\mathbf{r}, k) \approx \frac{\mu_0}{4\pi} \oint_{S_i} \frac{J_{i,l'}}{|\mathbf{r} - \mathbf{r}'|} dl' = \frac{\mu_0}{4\pi} \int_{v_i} \frac{J_{i,l'}(\mathbf{r}') \times (\mathbf{r} - \mathbf{r}')}{|\mathbf{r} - \mathbf{r}'|^3} dv' \tag{S14}$$

Discretizing the neuron into N compartments results in:

$$\mathbf{B}(\mathbf{r}, t) = \frac{\mu_0}{4\pi} \sum_{n=1}^{N} \int_{L_n} \frac{I_{axial}^n(t) \mathbf{dl} \times (\mathbf{r} - \mathbf{r_n})}{|\mathbf{r} - \mathbf{r_n}|^3} \tag{S15}$$

where $I_{axial}^n$ is the intracellular axial current for compartment n. $I_{axial}^n$ is determined by the potential difference $V_m^n - V_m^{n-1}$ between the membrane nodes which enclose the compartment:

$$I_{axial}^n = J_{i,l'}\pi a^2 = \frac{V_m^n - V_m^{n-1}}{\Delta s_n r_i^n} \tag{S16}$$

Integration of eq. (S15) along the center line of each compartment finally gives

$$\mathbf{B}(\mathbf{r},t) = \frac{\mu_0}{4\pi}\sum_{n=1}^{N}\frac{\mathbf{I}_{axial}^n(t) \times \hat{\rho}_n}{\rho_n}\left[\frac{h_n}{\sqrt{h_n^2 + \rho_n^2}} - \frac{l_n}{\sqrt{l_n^2 + \rho_n^2}}\right] \tag{S17}$$

where $\rho_n$ the distance perpendicular to the line compartment, $h_n$ the longitudinal distance from the end of the compartment, and $l_n = h_n - \Delta s_n$ the longitudinal distance from the start of the compartment.

## Comparison with the Analytical Solution for an Infinitely Long Cylinder

For the case of an infinite cylinder of radius $a$, an analytical solution can be derived for eq. (S11). As described above, the first term on the right hand side of eq. (S11) can be neglected, so that we only consider the surface integrals of the second and third terms here. The biological current sources are confined to the membrane and both the intra- and extracellular regions satisfy Laplace's equation for the electric potential. The boundary conditions are given by:

$$\phi_i - \phi_e = V_m \tag{S18}$$

$$-\sigma_i \frac{\partial \phi_i}{\partial n} = -\sigma_e \frac{\partial \phi_e}{\partial n} = J_m \tag{S19}$$

It can be shown that in this case, the exact solution of the potential distribution for an infinitely long cylinder can be calculated using the Fourier transform [73] and is given by:

$$\phi_e(r,k) = \frac{K_0(|k|r)}{\alpha(|k|a)K_0(|k|a)}U_m(k) \tag{S20}$$

$$\phi_i(r,k) = \frac{I_0(|k|r)}{\beta(|k|a)I_0(|k|a)}U_m(k) \tag{S21}$$

where

$$\alpha(|k|a) = -\left[1 + \frac{\sigma_e K_1(|k|a)I_0(|k|a)}{\sigma_i K_0(|k|a)I_1(|k|a)}\right] \tag{S22}$$

$$\beta(|k|a) = \left[1 + \frac{\sigma_i K_0(|k|a)I_1(|k|a)}{\sigma_e K_1(|k|a)I_0(|k|a)}\right] \tag{S23}$$

$I_0, I_1, K_0, K_1$ are modified Bessel functions and $U_m(k)$ is the Fourier transform of the transmembrane potential defined as

$$U_m(k) = \int_{-\infty}^{\infty} V_m(l') e^{jkl'} dl' \tag{S24}$$

The electric potentials $\phi_i$ and $\phi_e$ can be entered into eq. (S11) to calculate B in the frequency domain. When split according to the contribution of the intra- and extracellular axial surface currents, the magnetic fields are [8]:

$$B_i(r,k) = i\mu_0 a \frac{\sigma_i k}{\beta(|k|a)} K_1(|k|r) I_1(|k|a) U_m(k) \tag{S25}$$

$$B_e(r,k) = -i\mu_0 a \frac{\sigma_e k}{\alpha(|k|a)} K_1(|k|r) I_1(|k|a) U_m(k) \tag{S26}$$

We opted for the approach outlined above to calculate the magnetic field in the spatial frequency domain. In contrast to an alternative method that works in the spatial domain using an approximation of the membrane potential by Gaussian functions [6], it is faster and more accurate, as it does not suffer from fitting errors.

First, we use eq. (S25) and (S26) to compare the contributions of the external and internal currents to the magnetic field. For that, we apply the analytical expressions given in [7] to determine the membrane potential distribution, assuming a single axon with the following properties [9]:

| radius a | thickness $t_m$ | internal conductivity $\sigma_i$ | membrane conductivity $\sigma_m$ |
|---|---|---|---|
| $6 \times 10^{-5}$ m | $13.7 \times 10^{-9}$ m | $1\ \Omega^{-1} m^{-1}$ | $10^{-5}\ \Omega^{-1} m^{-1}$ |
| length L | velocity u | external conductivity $\sigma_e$ | membrane permittivity $\epsilon_m$ |
| 200 mm | $10\ \text{ms}^{-1}$ | $5\ \Omega^{-1} m^{-1}$ | $6.195 \times 10^{-12} C^2 N^{-1} m^2$ |

An action potential is induced at the left end of the axon, and a snapshot of the magnetic field distribution along the axon is taken at an (arbitrarily chosen) time point of 0.6 ms at which the action potential had travelled 6 mm. Fig. S2A&B show the strength of the magnetic field due to the internal and external current flow at a distance of 100 nm to the surface of the axon. The contribution of the external flow to the magnetic field is two orders of magnitude lower than that of the internal currents, confirming the above assumption used to derive eq. (S14). Specifically, the ratio between $B_i$ and $B_e$ is inversely proportional to the cell radius [8], rendering the contribution of $B_e$ to the total field insignificant for most neural cell radii.

Second, in order to validate the forward modeling scheme stated in eq. (S17), we simulate the membrane potential of a long straight axon in NEURON. The axon has active Hodgkin-Huxley channel dynamics and an intracellular resistance of $R_a = 100\ \Omega \text{cm}$ (i.e. $\sigma_i = 1/\Omega m$). Its length is 5 cm, and it is discretized in compartments with lengths smaller than the electrotonic space constant. The simulated membrane potential is used to derive the magnetic field at a distance of 100 nm to the surface of the axon, applying both the forward modeling scheme of eq. (S17) and the analytical approach of eq. (S25). The good overlap of both solutions shown in Fig. S2C&D confirms the derived forward modeling scheme.

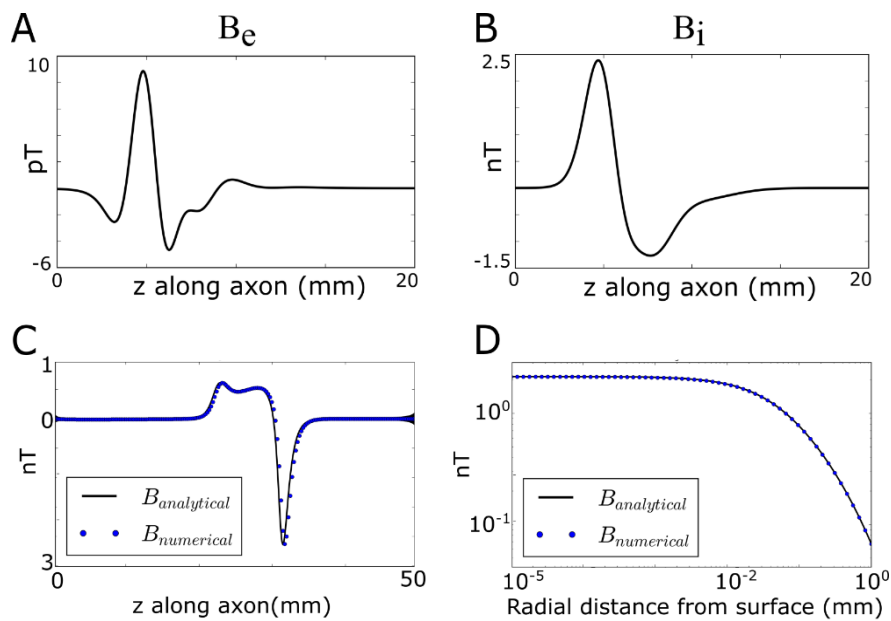

**Figure S2:** Analytically calculated and numerically simulated magnetic fields of a single axon. **(A)** Magnetic field component caused by the external current distribution (analytically determined), read out at a distance of 100 nm to the cell surface. **(B)** Magnetic field component caused by the internal current distribution (analytically determined, same distance as in A). **(C)** Comparison of numerically and analytically determined magnetic field distributions at a distance of 100 nm distance to the cell surface. The used membrane dynamics differ to those used in B, explaining the slightly different shape (please see main text for details). **(D)** Dependence of the peak magnetic field strength on the distance to the cell surface, determined both numerically and analytically.

## Comparison with Existing Studies: Single Axon case

We modeled a single axon to compare it with previously reported magnetometry measurements of axonal action potentials [10]. In that study, NV magnetometry measurements were performed for excised giant axons of two species, namely *Myxicola infundibulum* (worm) and *Loligo pealeii* (squid). In addition, recordings from axonal activity in intact worms was performed. Here, we compare our simulations with their results for the excised worm axon and the intact worm. In their first experiment, the giant axon of the worm was excised and measured at a temperature of 21°C and a standoff distance of 0.3 mm to the nerve center. In their second experiment, the worm was kept at 10 °C and the measurements were taken at a distance 1.2 mm from the nerve center. To compare our numerical methods with the reported measurement results, we modeled single axons with $\sigma_i = 1.5/\Omega m$ and diameters ranging between 200 μm and 400 μm in NEURON and estimated the resulting magnetic field based on the forward modeling scheme outlined above. The assumed temperatures were either 21°C (excised axon) or 10°C (axon of living worm). In the excised axon (Fig. S3A), the simulated peak magnetic field strength increases from around 1 to 3.5 nT with increasing diameter, which overlaps well with the measured field strengths. For the axonal activity of the living worm (Fig. S3B), the peak magnetic field strength decreases to one fourth of the one observed for the excised axon due to the increased distance. In addition, its duration is longer due to the lower ambient temperature that slows down the action potential. Our simulation results are in good agreement with the previously presented measurements [10], which show that the peak field strength was around 0.4 nT for a diameter of 300 μm.

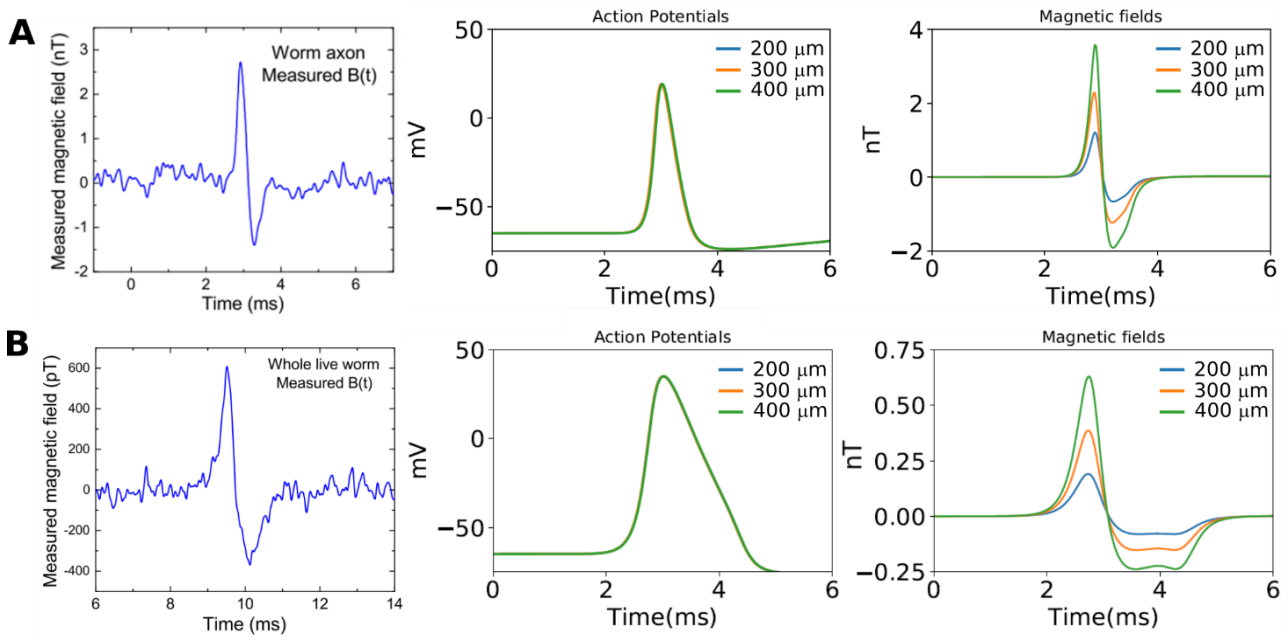

**Figure S3:** Single point measurement and simulations of the magnetic field. **(A)** Excised single giant axon at 21°C. Left: Measured magnetic field at distance of 300 μm from the center of the nerve (reproduced from [10]). Middle: Simulated action potential for different axon diameter. Right: Calculated total magnetic field at a distance of 300 μm. **(B)** Axonal activity of a living worm at 10 °C. Left: Measured magnetic field at distance of 1.2 mm from center of nerve (reproduced from [10]). Middle: Simulated action potential for different axon diameters. Right: Calculated total magnetic field at distance of 1.2 mm.

## Neural cell dynamics of the CA1 pyramidal cells

The CA1 pyramidal cells of the third scenario are modelled using previously reported morphology and biophysical properties [11]. The model files are available for public download under the ModelDB section of the Senselab database (http://senselab.med.yale.edu). The passive elements of the model comprise the cell membrane with a capacitance of $C_m = 1 \ \mu F/cm$ and a resistance of $R_m = 28 \ k\Omega cm^2$, and the intracellular space with a resistance of $R_a = 150 \ \Omega cm$. The active models elements comprise voltage-gated sodium channels and different potassium membrane currents (A-type and delayed rectifiers), with further details outlined next. Electrical stimulation of the Schaffer collaterals results in a temporally synchronous synaptic excitation of the CA1 pyramidal cells that are targeted by the collaterals. Here, this is simulated by excitation of the AMPA synapses of the pyramidal cells, which are situated on the apical dendrites (the stratum radiatum, S.R.) and the basal dendrites (the stratum oriens, S.O.). The synaptic events are modelled by time-dependent changes of the conductance of the synapses. These conductance changes are described by a double-exponential function $g_{syn} = \bar{g}_{syn} A^{-1} [e^{-t/\tau_{rise}} - e^{-t/\tau_{decay}}]$, with constant $\tau_{rise} = 1.5$ ms controlling the rise time, constant $\tau_{decay} = 2.5$ ms controlling the decay time and $\bar{g}_{syn}$ denoting the peak conductance[12]. Constant A is a normalization constant to set the maximum value of $g_{syn}$ to be $\bar{g}_{syn}$. These synaptic events inject currents of $I_{syn} = g_{syn}(V_m - V_{rev})$ at the membrane positions of the synapses (a reverse potential of $V_{rev} = 0$ mV is used). The simulations of the neural dynamics are conducted in NEURON, using the backward-Euler integration method to solve the cable equation with a fixed time step of 25 μs. For that, each section of a CA1 pyramidal cell (such as a dendritic branch) is subdivided into multiple cylindrical compartments to achieve compartment lengths smaller than the electrotonic space constant λ. Furthermore, each section is forced to have at least three compartments to be able to calculate axial currents from the membrane voltages as input to the above forward modelling scheme. The resulting models of CA1 pyramidal cells consist of 1427 compartments.

The stimulation of a CA1 pyramidal cell via the Schaffer collaterals is mimicked by synaptic events, which are generated at 40 randomly selected locations at the s. oriens and 40 random positions at the s. radiatum[13]. The time points of the synaptic events are modelled by truncated random functions with a normal distribution as follows:

1. Generate an excitation time $T_{exc}$

$$T_{exc} = \left(\frac{T_{start} + T_{stop}}{2}\right) + \left(\frac{T_{stop} - T_{start}}{4}\right) \mathcal{N}(0, \sigma)$$

2. If $T_{start} \leq T_{exc} \leq T_{stop}$, assign $T_{exc}$ as time point of the synaptic event, otherwise regenerate $T_{exc}$ until it remains in the desired time window.

$\mathcal{N}(0, \sigma)$ is a normal distribution with zero mean and standard deviation $\sigma$ (σ = 0.25, unless indicated differently), and represents the time jitter in the excitatory inputs. The first wave of synaptic events is generated using $T_{start} = 0$ and $T_{stop} = 25$ ms, and a second wave is generated using $T_{start} = 25$ and $T_{stop} = 50$ ms in order to mimic repeated electric stimulation at 40 Hz. The synaptic strength is controlled by setting the peak conductance $\bar{g}_{syn}$. In order to evaluate an upper limit for the magnetic field strengths that can occur merely due to EPSPs without inducing action potentials, a value of 0.3 nS was chosen. Consistent spiking activity could be created by selecting a strength of 0.6 nS (or higher, if indicated). It is worth noting that the number of modelled excitatory synaptic inputs is far lower than occurring for real neurons. In order to reduce the computational complexity of the simulations, the effects of a high number of excitatory synapses is mimicked by increasing the synaptic strengths instead.

## Additional Supplementary Figures

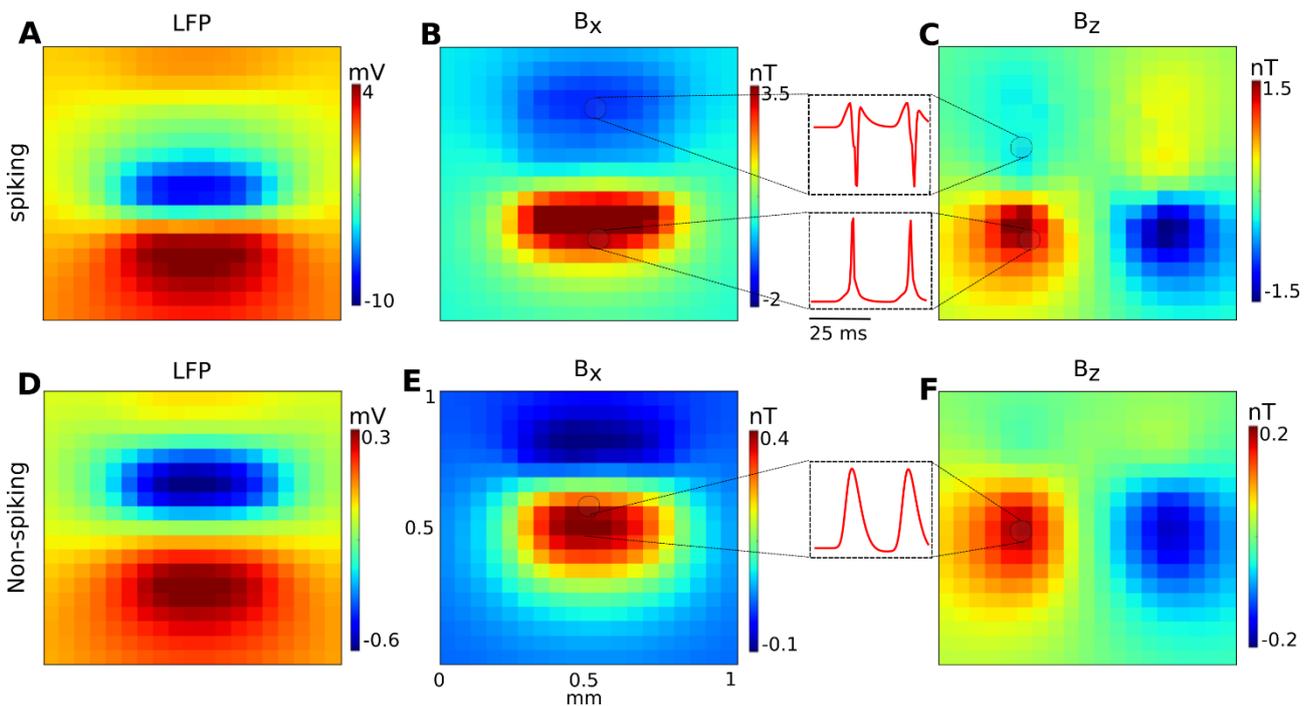

**Figure S4:** Simulation results showing the spatial distribution and the temporal shape of the extracellular fields for the CA1 subarea for the case that synaptic activity only occurs at the S.R. (**A**) LFP for the **spiking** case. (**B**) $B_X$ component of the magnetic field (spiking case). (**C**) $B_Z$ component of the magnetic field (spiking case). The **insets** depict the temporal shapes of the magnetic field components, extracted from the indicated positions (a time window of 50 ms is shown). The initial phases before the action potentials reflect the accumulation of EPSPs. (**D**) LFP for the **non-spiking** case. (**E**) $B_X$ component of the magnetic field (non-spiking case). (**F**) $B_Z$ component of the magnetic field (non-spiking case). The temporal shapes depicted in the insets reflects the accumulation of EPSPs.